# Intrinsic Geometric Analysis of the Network Reliability and Voltage Stability

N. Gupta, B. N. Tiwari, and S. Bellucci, Editorial Board Member,
Global Journal of Physics Express; and Journal of Physics & Astronomy

*Abstract-* **This paper presents the intrinsic geometric model for the solution of power system planning and its operation. This problem is large-scale and nonlinear, in general. Thus, we have developed the intrinsic geometric model for the network reliability and voltage stability, and examined it for the IEEE 5 bus system. The robustness of the proposed model is illustrated by introducing variations of the network parameters. Exact analytical results show the accuracy as well as the efficiency of the proposed solution technique.**

## I. INTRODUCTION

THE goal of this research paper is to advance the state-of-art in the power system reliability [1] and voltage stability [2], [3]. It is well known that for effective power system planning, the appropriate reactive compensation [4] is essential with a set of proper network parameters (resistance (R) and reactance (X) and planning [5], [6]. Subsequently, such a planning facilitates to reduce apparent power, by saving reactive power loss on the network.

Noticeably, in the varying reactive load scenario [7] [8] where the stability of voltage profile is necessary, the addition of the reactive components [9] of a desired bus or line has a significant from the perspective network designing [10]-[13]. As the result, it is obligatory to choose a finite value of the capacitance and reactor to keep voltage in the stability limit. The value of such parameters leads to design an interesting set of control power flow protection devices, i.e. tap changer, controlled switched compensator [14], [15]. Compensation strategy is used to improve voltage profile and power factor correction in AC transmission lines. Notice that the concept of reliability and voltage stability is related to the power quality issues [16] [17]. Most of the power quality issues are resolved by appropriate reactive power [18], which are controlled by capacitor bank [19] [20]. Such a scheme maintains the system performance, with almost stable system voltage profile and efficiency of the transmission line by improving impedance angle, steady-state controls and avoids blackouts.

From the perspective of power factor correction and highly disturbed bus (gradual or abrupt change of reactive load), an appropriate selection of the stabilizing compensation is essential [21], [22] [23]. With increasing complexity of the power system, the available slight linear methods yield a slow convergence (even non-convergence, as well). This follows from the fact that the linear analysis faces a number of nonrealistic implementation due to the uncertain behavior of load and contingency of the transmission lines, and thus the necessity of the restructuring.

Thus, our intrinsic geometric technology offers a self-consistent non-linear optimal solution and provides a resolution towards such disadvantages. In addition, there are myriads of methodologies and mathematical models for the selection of network parameters and compensation in the power system planning and operations, e.g., optimization techniques, genetic algorithm (GA), trial and error methodologies, fuzzy integrated with dynamic programming, artificial neural network (ANN) and heuristic analysis [24], [25] [26] [27] which have been in use in the most of the existing software's [28] [29]. It is worth emphasizing that the precise estimation of the transmission line parameters and associated compensation could hinder an occurrence of blackout in the power system with an increasing efficiency.

Up to the linear approximation, the prior solutions give a set of premature prominent results. This stipulates the need for an incorporation of the underlying non-linear effects. In this concern, our intrinsic geometric proposition yields an ample room to improve the network performance by the stable voltage profile of the power system. The previously revealed methodologies yield a set of approximate solutions, which are mostly iteratively accomplished. Such an implementation is generically realized by linearizing the corresponding flow equations, in a limited domain. Based on the above class of methodologies, the designed compensations are inefficient as per the need of modern society, due to slow convergence.

The present geometric model is a bootstrap towards these notions and indeed it is capable to provide strategic planning criteria for the most effective use of the network and the issue of network reliability. Form the very out-set of our geometric proposition, the voltage stability of the network and safety modes are an immediate consequence of the admissible network parameterization. With the help of correlation techniques [30]-[33] and critical point of an arbitrary network, the set of appropriate parameters and components can be identified for any finite component network. Interestingly, the global quantities of our model provide required set of safety alerts for the owner as well as to the regulator of the network. For a given total (complex) power, the proposed model is described as the subsequent of the innovation. The overall

N. Gupta is with the Department of Electrical Engineering, IITK, Kanpur, UP 208016, India, (e-mail: ngtaj@iitk.ac.in).
B. N. Tiwari is with the INFN-Laboratori Nazionali di Frascati, Via E. Fermi, 40 -- I-00044 Frascati (Rome), Italy (e-mail: tiwari@infn.lnf.it).
S. Bellucee is with the INFN-Laboratori Nazionali di Frascati, Via E. Fermi, 40 -- I-00044 Frascati (Rome) (e-mail: stefano.bellucci@lnf.infn.it).

methodology is implemented on the IEEE 5-bus system and the results are demonstrated from the expected feature of the proposed geometric model. The voltage levels of all the buses are assumed equal. This condition holds by the concept of the reverse engineering; where the results obtained show that the planning could be based on the calculated parameters, which keep the power system at a fixed voltage profile about which the network fluctuations are analyzed.

The rest of the paper is organized as follows. The present section summarizes the aforementioned techniques. In order to carry out this investigation, section-II provides the formulation of the problem. Section-III describes the details of the innovation and shows how it works in general. Section-IV evaluates the test of the network reliability and voltage stability and as the result proves the proposed work accuracy. The specific remarks of the present investigation are enlisted in Section-V. Finally, in Section-VI we draw conclusions and sketch the outlook for future investigations.

## II. Prior Mathematical Model Specification

In this section, to eliminate the effect of voltage fluctuation the optimum values of required L and C are determined. Also, we shall set up the method to predict the exact value of R and L in order to increase the reliability and efficiency of the network. From the described criteria, one can decide for single circuit or double circuit line between two buses to ensure the reliability of the network. All prior conventional solutions use load flow equations and characteristics and performance equations of transmission lines [34] [35]. In general, the power conservation equations associated with the real (resistive) and imaginary (reactive) branch parameters are respectively given by

$$P_i = \sum |V_i||V_j||Y_{ij}|\cos(\theta_{ij} + \delta_j - \delta_i) \quad (1)$$

$$Q_i = \sum |V_i||V_j||Y_{ij}|\sin(\theta_{ij} + \delta_j - \delta_i) \quad (2)$$

In the above equations, the phases are defined as: $\tan \theta_{ij} = \frac{X_{L_{ij}} - X_{C_{ij}}}{r_{ij}}$. Subsequently, we define impedance: $Y_{ij} = \frac{1}{Z_{ij}} = \frac{1}{r_{ij} + j(X_{L_{ij}} - X_{C_{ij}})}$ and phase $\delta_j = V_j/|V_j|$.

The steady state condition $|V_i| = 1$ makes the $i^{th}$ bus of the configuration in equilibrium under voltage fluctuations. The respective cases of our interest reduce to the following standard network considerations:
(i) For the real power flow with the considered resistances $\{r_i\}$, the $\theta_{ij}$ are defined by $\tan \theta_{ij} := (X_{L_{ij}})/r_{ij}$
(ii) In general, when the reactive power flow is allowed to possess a non zero voltage fluctuation, the $\theta_{ij}$ are defined as $\tan \theta_{ij} = \frac{X_{L_{ij}} - X_{C_{ij}}}{r_{ij}}$.

At this juncture, we propose that the local variations of the load are defined as the Hessian matrix of the total effective power in the network or at the chosen component. The robustness of the proposed model is further illustrated by introducing variations in the chosen circuit parameters and load. We share with the fact that the power fluctuations can be described by the critical exponent of the correlation equations. This model is able to predict the condition and state of the every branch of the network and thus whether it is robust from the perspective of power system planning.

## III. Proposed Methodology

The notion of power flow and framework of the Riemannian geometry supports the proposal of an intrinsic network reliability [36] [37] and global voltage stability [38], in case of abrupt load change at any disturbed bus of the power system. To begin with the novelty of the present advancement, we propose an admissible characterization of the network variables.

### A. Proposition

The impedances variables of the power network form an admissible basis.

*Proof:*
By performing a coordinate transformation, we demonstrate that the above proposition holds. To do so, let L, R, C be a set of mutually independent effective parameters of the network. Let us consider the following coordinate transformations on the LCR configuration

$$(r, X_L, X_C) = \left(R, \omega L, \frac{1}{\omega C}\right) \quad (3)$$

Thus, we notice that the Jacobian of the transformation is given by

$$J = \begin{bmatrix} 1 & 0 & 0 \\ 0 & \omega & 0 \\ 0 & 0 & -\frac{1}{\omega C^2} \end{bmatrix} \quad (4)$$

We observe that the determinant of the matrix J is $|J| = -\frac{1}{C^2}$. This provides a sort of fine tuning for the voltage fluctuations. This shows that the voltage fluctuations problem can be solved by a variable capacitor along with the variations of the other parameters. Consequently, we can analyze LCR fluctuations either in the impedance basis or in the basic component parameter basis, as long as the fluctuating component has a nonzero capacitor, viz., C≠0. Equivalently, both the above characterizations describe the same configuration of chosen component(s) and thus the whole network. For a nonzero base frequency, it is worth mentioning that the proposition holds for the LR component. This follows directly from the fact that the determinant of the Jacobian takes the value $|J| = \omega$. Some silent features of our network characterizations are the following:

For the power network system, our model specification yields that there exists a set of constant voltages in the equilibrium. Therefore, the corresponding phases between the reactive components are given by the extrema of the total power. The basis vectors are represented by the points on the intrinsic state-space manifold [30]-[33].

Consideration of the theory of random variation, along with the laws of equilibrium circuit configuration, $\{x_i\}$ leads to the Riemannian geometric structures [33]. Henceforth, over an equilibrium basis, defined as a finite set of the parameters $x_i=\{X_{L_i}, X_{C_i}, r_i\}$, they form the coordinate charts on the reactive configuration. Herewith, the states of interest are understood as the finite collection of points $\{x_i\}$.

Following [33], the present analysis attributes that the invariant distance between two arbitrary equilibrium states is inversely proportional to the random variation connecting the two states. In particular, the "less probable variation" signifies that the "states are far apart".

### B. Admissible Choice of Component(s)

In this model, our analysis utilizes the knowledge of foregoing real and complex power flow equations. Thereby, we determine the nature of the local and global reliability under the thermal limit of the network and voltage stability for uncertain voltage and thus fluctuations at the disturbed buses. By parallel lines, we predict the required augmentation for the network reliability and required value of C to keep the voltage level under stability limit in the current scenarios of power flow considerations. As the consequences of our proposed model, one determines the intrinsic geometric nature of the power system for a given set of parameters.

### C. Theoretical Motivation

The model under consideration is built from the viewpoint of intrinsic power system planning and perspective operation in modern applications. According to the existing electrical network design, we perceive that the following cases are interesting from the viewpoint of the planning. To determine the reliability of the network, we consider the loss in power flow, which modifies the set of inductive variables to the set of the combination of the resistances and inductances $\{r_i + X_{L_i}\}$. The voltage fluctuation problem is investigated by introducing the capacitances to the above set of variables $\{r_i + X_{L_i} + X_{C_i}\}$. Thus, the sets: $\{r_i + X_{L_i}\}$ and $\{r_i + X_{L_i} + X_{C_i}\}$ form basis vectors for the analysis of network reliability and voltage stability.

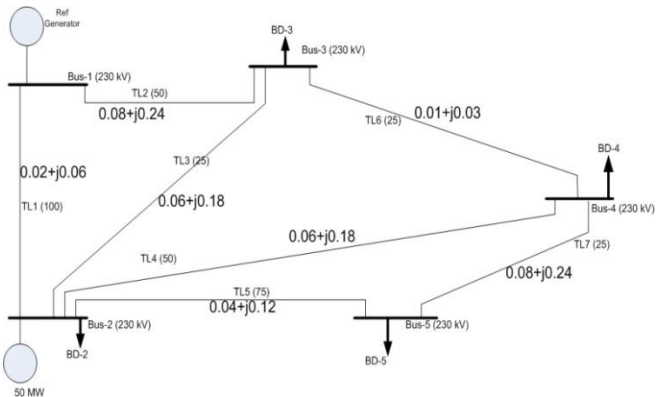

Fig.1. IEEE 5-Bus Circuit

### D. Planning Circuit

In this innovation, the utilities of the proposed model, network reliability and voltage stability of the power network are described for the circuit as for example shown in the Fig. 1. This model can be applied for a large class of high voltage networks, as well. The line parameters $(r_i + X_{L_i})$ are given on the respective transmission lines.

### E. Proposed Geometric Model

Herewith, we notice that the nearest neighbor pure pair correlations decay as the inverse square of the chosen components; whilst; the mixed pair correlations decay as the inverse of the chosen components [15]-[20]. The specific motivation of the present interest is to investigate the electrical configurations, when one wants to study the issue of network reliability and voltage stability of the circuit. The respective limiting network reliability of an arbitrary transmission line and limiting voltage stability of arbitrary component are effectuated with a fixed set of resistances, inductances and reactive capacitances.

- The metric tensor on the reactive state-space manifold is defined as:

$$g_{ij} = \partial_i \partial_j S_n(r, X_L, X_C) \qquad (5)$$

The above metric tensor may be further supported by the standard Taylor expansion of the effective (complex) power librated in the network.

- From the above consideration, it follows that a general network reduces to the power fluctuations on the reactive configuration, involving the resistances and reactance in the case of network reliability, or both the inductance and capacitances in the case of the voltage stability. In both the above cases, the parametric fluctuations yield the standard basis for the intrinsic reactive Riemannian manifold.

- The intrinsic scalar curvature of the reactive manifold is proportional to the correlation volume of the component configuration. This respectively signifies the global notions of the network reliability and voltage stability of the LR and LCR power systems.

- In fact, it turns out that the scalar curvature $R(x^i)$ scales as $R \propto \zeta^d(x^i)$, where $d$ is the spatial dimension of the underlying system and $\zeta(x^i)$ fixes the respective correlation scale for the reliability, and voltage stability of the chosen LR or RLC component of the network. A divergence of the scalar curvature indicates possible critical behavior or phase transitions in the underlying system, and at such a critical point the configuration is prone to have a breakdown for the current scenario of the power flow.

- The local and global network reliability and voltage stability of the electrical component(s) are thus taken into account in the unified manner with the inclusion of non-linear effects of the network fluctuations. Such variations can in principle be the ripple factor, current filtering and heating effects, which are simultaneously taken into account in the present analysis. This takes an intrinsic account against the linear transient analysis, and thus the present model gives a fast prediction for the integrated behavior of the desired component(s) of the network.

- The first step is to examine the intrinsic Wienhold geometry as the fluctuations about an equilibrium fixed point $(X_L, r)$,

for the chosen LR-component of specific transmission lines connected to disturbed bus.
- As the function of $X_L$ and $r$, the Wienhold metric tensor of the reactive space, defined as the maxima of the effective power, is given by

$$g_{ij} = \begin{bmatrix} \frac{\partial^2 S}{\partial X_L^2} & \frac{\partial^2 S}{\partial X_L \partial r} \\ \frac{\partial^2 S}{\partial X_L \partial r} & \frac{\partial^2 S}{\partial r^2} \end{bmatrix} \quad (6)$$

- The Riemann-Christoffel connections are defined as

$$\Gamma_{ijk} = g_{ij,k} + g_{ik,j} - g_{jk,i} \quad (7)$$

- In the case of the RL-circuit, the only non-zero Riemann tensor is

$$R_{X_L r X_L r} = \frac{N}{D} \quad (8)$$

where
$$N = S_{rr}(S_{X_L rr} S_{X_L X_L X_L} - S_{r X_L X_L}) + S_{X_L X_L}(S_{r X_L X_L} S_{rrr} - S_{X_L rr}) + S_{r X_L}(S_{r X_L X_L} S_{X_L rr} - S_{rrr} S_{X_L X_L X_L}) \quad (9)$$
and $D = 4(S_{rr} S_{X_L X_L} - S_{r X_L}^2) \quad (10)$

Here, the suffixes on the effective power S indicate the respective partial derivatives with respect to the network parameters.
- For the above type of two component networks, the Ricci scalar is given by:

$$R = \frac{2}{Det(g)} R_{X_L r X_L r} \quad (11)$$

- A novel aspect of our proposal is that its methodology allows one to predict both the reliability and the stability of the desired power system planning and its operation.
- For an arbitrary component finite network, the proposed intrinsic geometric formulation shows that the reliability corresponds to the hyper-space positivity, while the stability is determined by the positivity of the determinant of the state-space metric tensor.
- The global network reliability and global voltage stability are determined from the regularity properties of the scalar curvature of the underlying reactive-space invariants.

IV. TEST FOR AUTHENTICITY OF THE PROPOSED MODEL

Since the proposed model is the nonlinear improvement about the existing linear models, so, it is ensured that the intrinsic geometric analysis offers the best network reliability and voltage stability, over the nonlinear fluctuation of the real and complex power flow across the network component(s). Consequently, we examine the test of the reliability of transmission line by considering a LR component of the power flow. This follows from the consideration of the real power flow equation (1), without considering C compensation. Similarly, we demonstrate that the test of voltage stability of transmission line follows from the joint consideration of the real and complex power flow equations (1) and (2). In this apprehension, the specific possible considerations are given subsequently.

*Special Cases of Network Power Flow*

Our analysis offers that it is legitimate to analyze an evolution of the reactive component, infinitesimally fluctuating about an equilibrium configuration. The state of the art is then generically defined as the sum of the infinitesimal evolutions of an ensemble of the network components. This is equivalent to the consideration of the nearest neighbor interactions of the chosen component(s). For the foregoing consideration, we utilize the fact that a chosen component is in an equilibrium configuration, if the node voltages satisfy $|V_i|=1$.

A. *Test of Network Reliability*

In order to illustrate the real application of our intrinsic geometric proposition, let us consider the case of the RL-component. Without loss of generality, we can choose δ to be a constant in the equilibrium. Notice that the constant phases can be adjusted as per the requirement. This gives a liberty to choose to work either with the cosine angle or with the sine angle of the phase. Accordingly, the equation (2) shows that the power librated at the $i^{th}$ line is

$$P_i = Y_{i1} V_1 V_i \cos\left(\delta_i - \delta_1 - \tan^{-1}\left(\frac{X_{L_{i1}}}{r_{i1}}\right)\right) \quad (12)$$

With the standard trigonometric identity: $\cos\left(\tan^{-1}\left(\frac{X}{r}\right)\right) = \frac{r}{\sqrt{r^2+X^2}}$ and the fact that at the equilibrium $|V_i|=1$. In this case, the impedance of the component is defined as $|Y| = |Z|^{-1}$, and thus we find that the power librated at an arbitrary LR-component takes a simple expression.

Following the Eqn. (1), we see that the effective power across an arbitrarily LR component is given by

$$P(r,L) = \frac{r}{(r^2 + \omega^2 L^2)} \quad (13)$$

Thus, the components of the metric tensor on the inductive manifold are by

$$g_{rr} = 2\frac{r(r^2 - 3\omega^2 L^2)}{(r^2 + \omega^2 L^2)^3} ; g_{rL} = 2\frac{\omega^2 L(3r^2 - \omega^2 L^2)}{(r^2 + \omega^2 L^2)^3}$$
$$g_{LL} = -2\frac{r\omega^2(r^2 - 3\omega^2 L^2)}{(r^2 + \omega^2 L^2)^3} \quad (14)$$

- We observe that there is a $\overset{1}{\text{competition}}$ of reliability between the fluctuations of the resistance and inductance. For the typical transmission line, $r \ll X_L$, which is because a high value of $r$ posses high loss of the real power. The further losses due to the inductive component can be controlled by an appropriate compensation. Thus, from the fact that the fluctuation of r (due to uncertain real load) beyond a specified value of the inductance makes the inductive configuration unreliable.

- We observe that such an interesting intrinsic picture continues. In this case, we notice that the determinant of the metric tensor, viz., $Det(g)$, is given by

$$g(r,L) = -4\frac{\omega^2}{(r^2+\omega^2L^2)^3} \qquad (15)$$

- More explicitly, $g_{rr} > 0$ further shows that a relatively large resistive load is reliable, whereas, on the other hand, the condition $g_{LL} < 0$ implies that the inductive load is reliable, which is affected by the resistive loading on the system. Thus, the boundary of reliability to choose parameters are given by the following {r, L} curve

$$r^2 - 3\omega^2 L^2 = 0 \qquad (16)$$

Thus, the appropriate reactors can be designed for varying resistive demand and also to reduce resistance of the line, as required by the parallel lines can be chosen to keep the power flow of the network reliable.

- The negativity of the determinant of the metric tensor $Det(g)$ shows that the entire power system becomes unreliable, if both the resistive and the inductive components, viz., {r, L}, vary simultaneously. Thus, to further reduce the inductance effect of the transmission line(s) connected to a bus bar, the appropriate compensation is required which can be computed as given in next section.
- From the equation (11), we see that the Ricci scalar curvature vanishes identically for the RL metric tensor, as defined by the (14). The vanishing of the intrinsic scalar curvature demonstrates that the network, as the finite union of the LR-components, is free from global instabilities, and thus arbitrary LR-components remain globally reliable.

*B. Test of Voltage Stability*

With the notions introduced in the case of the network reliability, the tuning of the line parameter may be carried out by $r$ and $L$, and we may similarly add a non-zero capacitance to the LR-component, and thus get the resulting RLC-component to maintain some voltage profile. In this case, following the Eqns. (1) and (2), we find that the real and reactive power librated on the $i^{th}$ bus through $i=1$ line is

$$P_i = Y_{i1}V_1V_i\cos\left(\delta_i - \delta_1 - \tan^{-1}(\frac{X_{L_{i1}}-X_{C_{i1}}}{r_{i1}})\right) \qquad (17)$$

$$Q_i = Y_{i1}V_1V_i\sin\left(\delta_i - \delta_1 - \tan^{-1}(\frac{X_{L_{i1}}-X_{C_{i1}}}{r_{i1}})\right) \qquad (18)$$

At the local equilibrium, an appropriate choice of the load angles are determined by the choice: $\delta_i = \delta_1$. Thus, the unified power librated on the $i^{th}$ bus, due to the connected lines, reduces to the following formula

$$S(r, X_L, X_C) = \frac{(r + X_L - X_C)}{r^2 + (X_L - X_C)^2} \qquad (19)$$

From the above equation, a direct substitution of $X_L$ and $X_C$ shows that the total effective power through the LCR component takes the following form

$$S(r,L,C) = \frac{r + \omega L - \frac{1}{\omega C}}{r^2 + (\omega L - \frac{1}{\omega C})^2} \qquad (20)$$

As mentioned for the LR- component, the components of the metric tensor can be computed from the definition of the Hessian matrix: $Hess(P(r, L, C))$. Thus, we see that the components of the metric tensor, as the union of the real and imaginary power flow fluctuations, are given by

$$g_{rr} = \frac{2\omega^3 C^3 \begin{pmatrix} r^3\omega^3C^3 - 3r\omega^5C^3L^2 + 6r\omega^3C^2L \\ -3r\omega C + 3r^2\omega^4C^3L - 3r^2\omega^2C^2 \\ -\omega^6L^3C^3 + 3\omega^4L^2C^2 - 3\omega^2LC + 1 \end{pmatrix}}{(r^2\omega^2C^2 + \omega^4L^2C^2 - 2\omega^2LC + 1)^3}$$

$$g_{rL} = \frac{-2\omega^4 C^3 \begin{pmatrix} r^3\omega^3C^3 - 3r\omega^5C^3L^2 + 6r\omega^3C^2L \\ -3r\omega C - 3r^2\omega^4C^3L + 3r^2\omega^2C^2 \\ +\omega^6L^3C^3 - 3\omega^4L^2C^2 + 3\omega^2LC - 1 \end{pmatrix}}{(r^2\omega^2C^2 + \omega^4L^2C^2 - 2\omega^2LC + 1)^3}$$

$$g_{rC} = \frac{-2\omega^2 C \begin{pmatrix} r^3\omega^3C^3 - 3r\omega^5C^3L^2 + 6r\omega^3C^2L \\ -3r\omega C - 3r^2\omega^4C^3L + 3r^2\omega^2C^2 \\ +\omega^6L^3C^3 - 3\omega^4L^2C^2 + 3\omega^2LC - 1 \end{pmatrix}}{(r^2\omega^2C^2 + \omega^4L^2C^2 - 2\omega^2LC + 1)^3}$$

$$g_{LL} = \frac{-2\omega^3 C \begin{pmatrix} r^3\omega^3C^3 - 3r\omega^5C^3L^2 + 6r\omega^3C^2L \\ -3r\omega C + 3r^2\omega^4C^3L - 3r^2\omega^2C^2 \\ -\omega^6L^3C^3 + 3\omega^4L^2C^2 - 3\omega^2LC + 1 \end{pmatrix}}{(r^2\omega^2C^2 + \omega^4L^2C^2 - 2\omega^2LC + 1)^3}$$

$$g_{LC} = \frac{-2\omega^3 C \begin{pmatrix} r^3\omega^3C^3 - 3r\omega^5C^3L^2 + 6r\omega^3C^2L \\ -3r\omega C + 3r^2\omega^4C^3L - 3r^2\omega^2C^2 \\ -\omega^6L^3C^3 + 3\omega^4L^2C^2 - 3\omega^2LC + 1 \end{pmatrix}}{(r^2\omega^2C^2 + \omega^4L^2C^2 - 2\omega^2LC + 1)^3}$$

$$g_{CC} = \frac{-2\omega^2 \begin{pmatrix} \omega L - r - 3r^2\omega C + 3\omega^5C^2L^3 + 3r^3\omega^2C^2 \\ +3r^2\omega^3C^2L + 3r\omega^4C^2L^2 - 2r\omega^6L^3C^3 \\ -2r^3\omega^4C^3L - 3\omega^3L^2C - \omega^7C^3L^4 + r^4\omega^3C^3 \end{pmatrix}}{(r^2\omega^2C^2 + \omega^4L^2C^2 - 2\omega^2LC + 1)^3}$$

(21)

From the viewpoint of the voltage stability of the *i*-bus, the following issues are important.

*B.1. Surface-Stability*

- The principle minor $P_2$, defining the stability of *LC*-surface variations is depicted in Fig. 2, with respect to inductance $L$ and capacitance $C$ for $r = 0$. It shows that an appropriate choice of $C$ is required for relatively smaller value of $L$. For instance, $C$ should be smaller than 0.5 p.u.. Furthermore, it turns out that the deviation of the complex power grows rapidly, e.g. like $e^{22}$, to a large negative value. The general expression of the minor is not very elegant to present in this paper. Specifically, for $r = 0$, we find that the surface minor of the *LC*-oscillations becomes unstable with the following value for the limiting minor

$$P_2 = \frac{-4\omega^6 C^4 \begin{pmatrix} 1-20\omega^6 L^3 C^3 - 20\omega^8 L^3 C^5 + 15\omega^{10} L^4 C^6 \\ +15\omega^8 L^4 C^4 - 6\omega^4 C^3 L - 6\omega^{10} L^5 C^5 - 6\omega^2 LC \\ -6\omega^{12} L^5 C^7 + 15\omega^4 L^2 C^2 + 15\omega^6 L^2 C^4 \\ +\omega^{14} L^6 C^8 + \omega^{12} L^6 C^6 + \omega^2 C^2 \end{pmatrix}}{(1+\omega^4 L^2 C^2 - 2\omega^2 LC)^6} \quad (22)$$

### B.2. Volume Stability

The determinant of the metric tensor defining voltage stability of the LRC-variations has been depicted in the Fig. 3. The determinant of the metric tensor is plotted with respect to inductance $L$ and capacitance $C$ for $r = 0$. It shows that for relatively smaller values of $L$ the corresponding value of the $C$ should be less than $0.5$. Both the surface and volume stability of the *LCR* oscillations thus remain the same for the zero value of the resistance. In a range of $\{L, C, r=0\}$, it is worth mentioning that the qualitative feature of the determinant of the metric tensor remains the same as for the surface minor. In the limit of $L = 0$ and $r = 0$, we notice that the determinant of the metric tensor reduces to the following expression

$$g = 8(1 - 3\omega^2 C^2)\omega^7 C^3 \quad (23)$$

- As we move to a larger value of $C$ with respect to the L, we observe that the deviated complex power fluctuations grow rapidly and reach to a very high value. In contrast to the case of the network reliability, the deviation in the voltage stability grows to the order of $e^{25}$. For the AC frequency of $f = 50$ Hz, we notice further that the limiting $L = 0, r = 0$ determinant of the metric tensor remains positive for a large $C$. Specifically, our analysis promises prominently stable choices for the values of $L$ and $C$. Such a characterization offers required regulation for the power supply.

### B.3. Global Stability

The state-space global stability of the LCR network fluctuations is depicted in the Fig. 4, with respect to the inductance $L$ and capacitance $C$ for the choice of $r = 0$. It shows that the global nature of limiting LCR oscillations is acceptable for a range of the capacitance. In the limit of the zero capacitance, the system shows a large global correlation of the order $e^3$. For the small L, e.g. $L < 0.4$, we find that such a LCR component is globally weakly correlated. However, for a relatively larger value of the $L$, the component demands a larger value of the $C$.

As we move to a larger value of the $C$, we observe that the deviation of the total power ceases rapidly and the whole component becomes globally uncorrelated. The general expression for the curvature scalar is rather intricate. Interestingly, for the choice of $L = 0$ and $r = 0$, we notice that the limiting curvature scalar takes the following form

$$R = \frac{1}{2}\left(\frac{-6 + 25\omega^2 C^2 - 51\omega^4 C^4}{(1 - 3\omega^2 C^2)^2 \omega C}\right) \quad (24)$$

- It is worth mentioning that the limiting system becomes unstable for the vanishing value of the determinant of the metric tensor. From Eqn. (23), we see that the scalar curvature of the limiting $L = 0, r = 0$ configuration diverges, and thus we find in this limit that the entire system becomes highly unstable. The identical notion holds for the general values of the *L, C* and *r*.

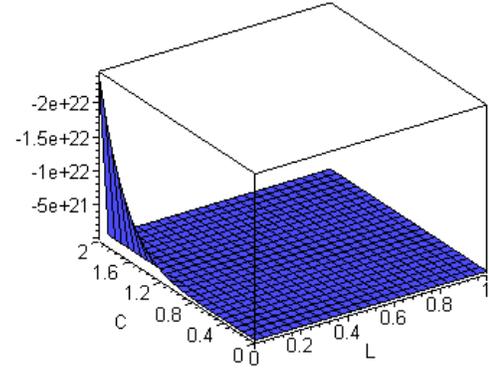

Fig.2. Surface Stability

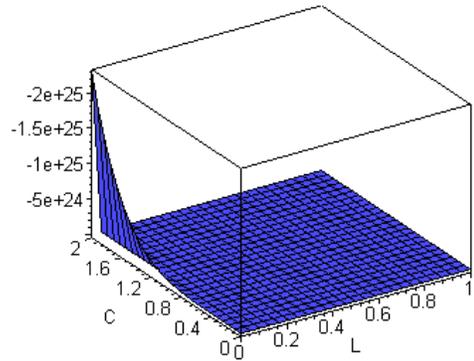

Fig. 3. Volume Stability

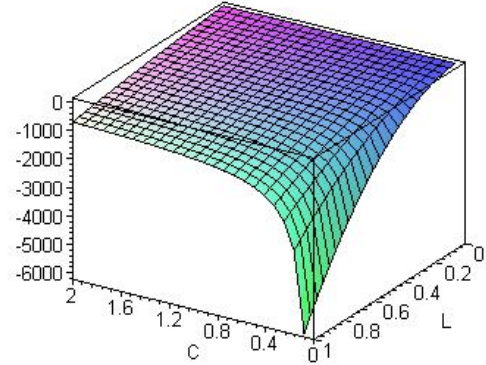

Fig.4. Global stability

Following the Fig. 2, Fig. 3 and Fig. 4), where the third dimension shows the minor, determinant of the metric tensor and global scalar curvature of the system, we may conclude for $r = 0$ that the most appropriate values of the capacitor should lie in the range $0.1 < C < 0.5$. For the stability of the LCR-component, this follows from the fact that the surface minor and determinant of the metric tensor deflate for $C > 0.5$ and $0.1 < C$, the scalar curvature blow ups. Thus, we can design an appropriate compensator and reactor to keep system voltage maintained.

## V. Specific Remarks

### A. Limiting voltage stability

The configuration with $r = 0$ corresponds to the limiting voltage stabilization, as shown in the Fig. 2, Fig. 3, Fig. 4. These diagrams describe the nature of the LC fluctuations of the LCR-component about fixed AC base line, e.g., a constant frequency of *50 Hz* or *60 Hz*.

### B. Limiting reliability

The case of $C \to 0$ corresponds to the limiting reliability of the LCR-component. This notion may be analyzed in a similar manner of $r \to 0$. Such considerations describe LR fluctuations of the LCR-component about an almost fixed DC base. For a variable base inductance, the inductive fluctuations introduce a non-zero ripple in the LR-component.

### C. Non-linear reliability and stability

As per the validity of the parameterization of LCR components, the Fig. 4 show that the zero capacitance is a singular point of the reactive configuration. This observation is in accordance with the regularity condition of the Jacobian matrix of the coordinate transformation of fluctuating LCR component. Herewith, we find that the strongly correlated reactive space notions remain valid even for the limiting zero resistance component $r \to 0$.

TABLE-I
QUANTITATIVE VERIFICATION OF THE RL COMPONENT

| T | r [pu] | L[pu] | Det(g) |
|---|--------|-------|--------|
| 1 | 0.02 | 0.60 | -852772.88 |
| 2 | 0.08 | 0.24 | -208.19 |
| 3 | 0.06 | 0.18 | -1169.78 |
| 4 | 0.06 | 0.68 | -0.41 |
| 5 | 0.04 | 0.12 | -13324.57 |
| 6 | 0.01 | 0.03 | -54577464.92 |
| 7 | 0.08 | 0.024 | -22377541.16 |

TABLE-II
QUANTITATIVE VERIFICATION OF THE RLC COMPONENT

| T | r[pu] | L[pu] | C[pu] | P2 | Det(g) | R |
|---|-------|-------|-------|-----|--------|---|
| 1 | 0.02 | 0.60 | 0.30 | -272.93 | -11519.51 | -4.92 |
| 2 | 0.08 | 0.24 | 0.025 | -0.20e-2 | 0.68 | 36.89 |
| 3 | 0.06 | 0.18 | 0.020 | -0.74e-3 | 0.27 | 46.58 |
| 4 | 0.06 | 0.68 | 0.020 | -0.14e-2 | 0.79 | 41.45 |
| 5 | 0.04 | 0.12 | 0.015 | -0.21e-3 | 0.96e-1 | 62.82 |
| 6 | 0.01 | 0.03 | 0.010 | -0.38e-4 | 0.24e-1 | 95.40 |
| 7 | 0.08 | 0.024 | 0.025 | -0.15e-2 | 0.39 | 39.90 |

T tends for transmission line

To check the reliability of the transmission line, we offer an explicit test for chosen component.

From the Table I, we notice that all the transmission lines have a negative determinant of metric tensor. This indicates that there is a need to strengthen the transmission line by satisfying (16) for a reliable power supply. For a given shunt capacitance, the corresponding test of the stability is depicted in the Table II. This table shows that power flow in the transmission lines demanding certain variation make the system unstable. Thus, there is need to provide compensation for the transmission lines, except the first line. In this case, this demonstrates the quantification of the Figs. 2, 3 and 4.

### D. Stability of Minimally Coupled Bus

From the above consideration, if there are *n*- minimally coupled transmission lines connected to a concerned bus, then the stability of the bus is defined by the metric tensor

$$A_{yz} = diag(g_{ij(n)}), \qquad (25)$$

where $A_{yz} = \partial_y \partial_z S(x^1, x^2, \ldots, x^n)$; and $y, z = 1, 2, \ldots, n$.

In the above setup, the elements appear as the diagonal elements of the total system. Thus, all the principle minors remain equal and greater than zero for a stable voltage profile. The determinant: $det(g_{ij(n)})$ dependents on the foregoing determinants $det(g_{ij(n-1)})$, for all n. This characterization improves the performance of the power system in situation of sudden load change or transmission lines outage redundancy.

## VI. Conclusion and Outlook

The present paper explores the issue of planning environment of the power industry. From the perspective of engineering applications, the issue of the planning and operation pertaining to the network reliability and bus voltage stability has been examined by the intrinsic geometric consideration. Specifically, we have considered an intrinsic geometric model for the limiting reliability and limiting voltage stability analysis of the electrical networks. The correlation model has been converted into the intrinsic geometric model, on order to encounter the non-linear effects of stochastic nature encoded in power systems with an appropriate optimization of the components. The robustness of the proposed model is illustrated by introducing variations of the circuit parameters. In this model, the perspectives of reliability and stability of a component are directly accomplished through the framework of intrinsic Riemannian geometry.

From the very definition of the present approach, we have offered a systematic exposition and conversion of the basic principles of covariance and correlation techniques unifying: *i*) intrinsic geometric model, *ii*) power network analysis advances, *iii*) system stability, and *iv*) high efficiency and performance. These considerations have led our proposition viable over the nonlinear effects. It is worth mentioning that the speed of the proposed analysis can be kept arbitrarily large for choosing compensation strategy, transient and power flows subject to a variation of the resistive and inductive loads. The power flows include the consequences arising from the real and imaginary power of an electrical circuit of finitely many nodes. The notion of the criticality is explicated from the determination of critical exponents of the global reliability and voltage stability. This investigation offers the network reliability and voltage stability of the buses under the local variation on the load impedance(s). Such a methodology can provide power industry to re-engineer the network design with an improved insurance of power systems.

In a nutshell, the proposed intrinsic geometric model is capable of solving the following specific problems; (i) global picture of the network reliability and bus bar voltage stability, (ii) non-linear effects of the network reliability and bus bar voltage stability and (iii) transient phenomenon. This model further supports the detailed analysis and studies pertaining to prevent risk of the large-scale blackouts. Based on the above considerations, we can efficiently maintain the underlying configuration economic and thus to keep the accessible power market operating. Furthermore, in accordance with the existing network planning strategies, our geometric proposition may provide suitable locations for constructing the generation plants. One can further speed up the operating capability of the network owners, in order to keep the desired network efficient with improving performance of the power system. The proposed methodology is effectively generous and it can pace the research pertaining to future optimal electricity market designs and planning. This problem is left for a future study.

## VII. Acknowledgments

BNT thanks Prof. V. Ravishankar for his support and encouragements and "INFN-Italy" for postdoctoral research fellowship towards the realization of this research. NG thanks Prof. Prem K. Kalra and Prof. R. Shekhar towards the basic motivation of this work.